\documentclass[11pt]{article}
\usepackage{fa2025}
\usepackage{amsmath}
\usepackage{cite}
\PassOptionsToPackage{hyphens}{url}

\usepackage{url}
\usepackage{graphicx}
\usepackage{color}
\usepackage{siunitx}
\usepackage[utf8]{inputenc}
\usepackage[none]{hyphenat}
\usepackage{hyperref}

\title{The Extended SONICOM HRTF Dataset and Spatial Audio Metrics toolbox}

\multauthor
{Katarina C. Poole$^{1*}$ \hspace{1cm} Julie Meyer$^{1,2}$ \hspace{1cm} Vincent Martin$^{1,3}$ \hspace{1cm} Rapolas Daugintis$^1$}
{\bfseries {Nils Marggraf-Turley$^1$ \hspace{1cm}  Jack Webb$^1$ \hspace{1cm} Ludovic Pirard$^1$ \hspace{1cm} Nicola La Magna$^1$}\\
{\bfseries {Oliver Turvey$^1$ \hspace{1cm} Lorenzo Picinali$^1$}}\\
$^1$ Dyson School of Design Engineering, Imperial College London, UK\\
$^2$ Department of Engineering, Kings College, London, UK\\
$^3$  ENTPE, Ecole Centrale de Lyon, CNRS, LTDS, UMR5513, France
\correspondingauthor{katarina.poole@imperial.ac.uk}{Poole et al.}
}

\begin{document}

\maketitle
\begin{abstract}
Headphone-based spatial audio uses head-related transfer functions (HRTFs) to simulate real-world acoustic environments. HRTFs are unique to everyone, due to personal morphology, shaping how sound waves interact with the body before reaching the eardrums. Here we present the extended SONICOM HRTF dataset which expands on the previous version released in 2023. The total number of measured subjects has now been increased to 300, with demographic information for a subset of the participants, providing context for the dataset’s population and relevance.  The dataset incorporates synthesised HRTFs for 200 of the 300 subjects, generated using Mesh2HRTF, alongside pre-processed 3D scans of the head and ears, optimised for HRTF synthesis. This rich dataset facilitates rapid and iterative optimisation of HRTF synthesis algorithms, allowing the automatic generation of large data. The optimised scans enable seamless morphological modifications, providing insights into how anatomical changes impact HRTFs, and the larger sample size enhances the effectiveness of machine learning approaches. To support analysis, we also introduce the Spatial Audio Metrics (SAM) Toolbox, a Python package designed for efficient analysis and visualisation of HRTF data, offering customisable tools for advanced research. Together, the extended dataset and toolbox offer a comprehensive resource for advancing personalised spatial audio research and development.
\end{abstract}

\section{Introduction}\label{sec:introduction}

Immersive technologies have come to the forefront of spatial audio research, driving the need for extensive datasets of Head-Related Transfer Functions (HRTFs). These datasets are crucial for personalised spatial audio rendering in applications such as virtual reality, augmented reality and hearing aid development. Several HRTF datasets have been created and made publicly available, using both real and artificial heads, allowing for reproducible research in spatial acoustics and auditory perception \cite{guezenocWideDatasetEar2020,algaziCIPICHRTFDatabase2001,carpentierMeasurementHeadRelatedTransfer2014,majdak3DLocalizationVirtual2010,watanabeDatasetHeadrelatedTransfer2014,bomhardtHighResolutionHeadRelatedTransfer2016,sridharDatabaseHeadRelatedTransfer2017,armstrongPerceptualEvaluationIndividual2018,yuNearfieldHeadrelatedTransferfunction2018,brinkmannCrossevaluatedDatabaseMeasured2019a,ghorbalComputedHRIRsEars2020}. Among these is the SONICOM HRTF dataset \cite{isaacSonicomHRTFDataset2023}, one of the largest HRTF datasets available, which in its initial release  included 200 HRTFs along with photogrammetry data and high resolution 3D scans of the heads of the subjects. The SONICOM dataset has been cited in 24 different studies at the time of writing and currently represents one of the more widely used publicly available HRTF datasets. 

Despite its contributions to the field, like many existing HRTF datasets, the SONICOM dataset faces limitations that reduce its applicability in modern spatial audio research. Firstly, the amount of available data remains insufficient to effectively train and validate deep learning models. While some efforts have been made to merge existing datasets by harmonising HRTFs \cite{geronazzoTechnicalReportSONICOM2024}, these attempts have been constrained by measurement inconsistencies across datasets \cite{pauwelsHartufoToolkitMachine2023}.  Secondly, traditional HRTF acquisition methods are resource-intensive and prone to errors. Measurements require an anechoic environment, a large loudspeaker array, and precise microphone placement, making the process costly and time-consuming. Additionally, small subject movements during measurement can introduce inaccuracies, further complicating dataset reliability.

One approach to reducing measurement complexity is HRTF up sampling, where data is interpolated from a limited number of measured source positions \cite{hoggHRTFUpsamplingGenerative2024}. However, this technique can introduce inaccuracies, limiting its effectiveness. An alternative solution is HRTF synthesis, where computational models simulate HRTFs based on detailed 3D scans of a subject’s morphology. While prior work has explored this method \cite{ziegelwanger2015mesh2hrtf,brinkmannRecentAdvancesOpen2023a,brinkmannCrossevaluatedDatabaseMeasured2019a}, our dataset represents a large-scale implementation of HRTF synthesis paired with measured HRTFs and the underlying modelling data. 

In this work we extend the SONICOM HRTF dataset with the addition of 100 more measured subjects incorporating associated pre-processed 3D scans and HRTFs simulated using Mesh2HRTF.  We also introduce a Python toolbox called Spatial Audio Metrics (SAM) aimed at HRTF analysis and visualisation. These tools were designed to mitigate the limitations of traditional measurements techniques while expanding the dataset’s scope for machine learning applications. By leveraging pre-processed 3D scans, we can investigate the efficacy and iteratively improve computational generated HRTFs. This extension not only enhances the dataset’s usability for spatial acoustics research, but also paves the way for scalable, cost-effective HRTF personalisation in immersive applications.

\section{Methods}\label{sec:methods}

To generate HRTFs based on anthropometric features, we employed 3D scans from the existing SONICOM dataset \cite{isaacSonicomHRTFDataset2023} as well as newly acquired scans from the extended dataset. However, not all scans could be processed due to various errors, particularly those with excessive distortions around the pinnae, rendering some scans unusable. The initial scanning process produced point clouds with a 0.5mm resolution, capturing the head and portions of the shoulders. HRTFs were then generated using Mesh2HRTF \cite{ziegelwanger2015mesh2hrtf}, a simulation framework based on the boundary element method (BEM). This method requires a watertight mesh with no holes, duplicate vertices, or intersecting faces, and with all faces facing outwards \cite{ziegelwanger2015mesh2hrtf,brinkmannRecentAdvancesOpen2023a,ziegelwanger2015numerical}. 
To meet these requirements, the raw point clouds were converted into watertight meshes using ExScan Pro Software, applying minimal filtering and smoothing, and interpolating missing regions to ensure a watertight mesh. To ensure the subject’s face was oriented forward, we aligned the scans to the Frankfurt plane. Additionally, all head and facial hair were removed, and the scans were truncated below the neck for consistency (see Figure 1A-B). To address inconsistencies across scans, we provide two versions of each processed scan. The pre-processed scan retains minimal modifications to maintain anatomical accuracy. The plugged scan has the ear canal occluded up to the entrance, as both HRTF measurements and simulations typically assume that the ear canal does not introduce direction-dependent effects (see Figure 1C-D). This ensures that simulations generated from the scan are taken at the entrance rather than within the canal itself.

\begin{figure}[ht]
 \centerline{
 \includegraphics[width=7.8cm]{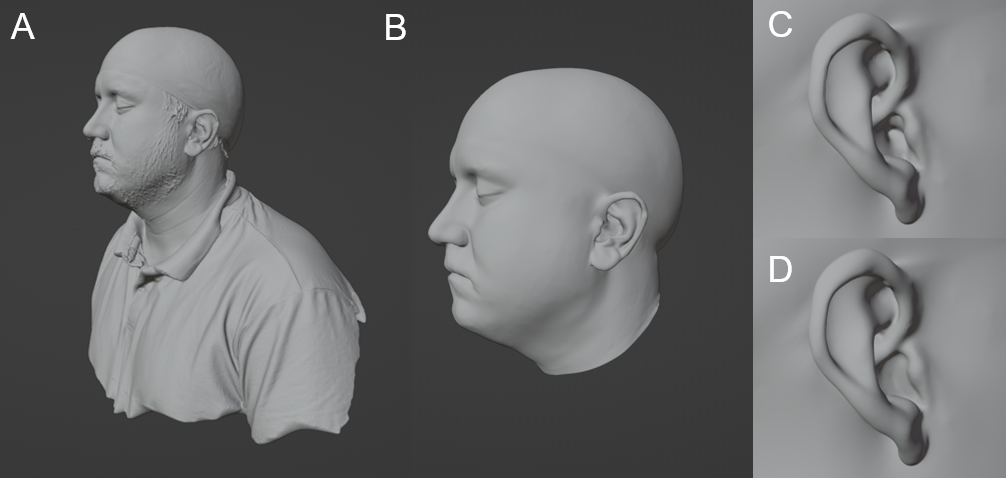}}
 \caption{Processing of 3D scans for HRTF simulation. (A) Raw 3D scan before processing (B) Processed scan, aligned to the Frankfurt plane with hair and extraneous features removed C) Preprocessed ear canal (D) Plugged ear canal variant.}
 \label{fig:example}
\end{figure}

Additionally, the scans were graded to optimise computational efficiency while preserving accuracy \cite{palmCurvatureadaptiveMeshGrading2021}. The resolution of the mesh was maintained around the ipsilateral pinna but reduced on the contralateral side to optimise processing. This approach was based on observations showing that decreasing the resolution of the mesh with increasing distance from the ear for which the HRTF is approximated can significantly reduce computation time \cite{palmCurvatureadaptiveMeshGrading2021}. The graded meshes were then used to simulate the HRTFs with Mesh2HRTF. HRTFs were simulated for frequencies between 0 Hz and 24 kHz in steps of 150 Hz for each position in the SONICOM measurement grid (793 positions, between -45° and 225° in elevation, 360° around the subject in 5° increments). 

\section{Description of Dataset and Toolbox}\label{sec:results}

\subsection{Extended SONICOM Dataset}\label{subsec:dataset}

The dataset includes 200 synthetic HRTFs and 300 measured HRTFs, making it one of the largest publicly available HRTF datasets. As with the previous release, it contains measured HRTFs, unprocessed 3D scans, and photogrammetry. For a full description of the existing data, please refer to \cite{isaacSonicomHRTFDataset2023}. The extended dataset introduces an additional SYNTHETIC\_HRTF folder within each PXXXX participant folder (where the scan quality was sufficient for HRTF generation). Here, PXXXX represents the participant number. The folder contains the following files:
\begin{itemize}\setlength\itemsep{-0.25em}
\item PXXXX\_preprocessed.stl
\item PXXXX\_plugged.stl
\item PXXXX\_graded\_left.stl
\item PXXXX\_graded\_right.stl
\item HRIR\_SONICOM\_44100.sofa
\item HRIR\_SONICOM\_48000.sofa
\end{itemize}

The pre-processed and plugged STL files represent the raw 3D scans with minimal modifications, while the graded meshes reduce computational complexity by optimising resolution based on ear proximity. These meshes were used to generate HRTFs via Mesh2HRTF, and the results are provided in SOFA file format \cite{majdakSpatiallyOrientedFormat2013} at sample rates of 44.1 kHz and 48 kHz. Additionally, demographic data is available as a CSV file, listing participant age (in years), ethnic group, and sex at birth for those who consented to share their information. The database is publicly available under the MIT license at: 
\href{https://transfer.ic.ac.uk:9090/#/2022_SONICOM-HRTF-DATASET}{https://transfer.ic.ac.uk:9090/\#/2022\_SONICOM- \break HRTF-DATASET/} 

\subsection{Spatial Audio Metrics Toolbox}

To support the numerical evaluation of HRTFs, we developed the Spatial Audio Metrics Python toolbox. This open-source framework was designed to provide a robust and extensible solution for analysing and comparing HRTF datasets. Our implementation allows users to compute key spatial audio parameters, including spectral distortion, interaural time and level differences (ITD/ILD), which are essential for assessing HRTF quality and consistency. The development of Spatial Audio Metrics was driven by the need for a standardised, reproducible evaluation methodology that could rapidly compare HRTFs numerically and visually from different subjects and synthesis methods, necessary for iteratively improving HRTF measurement and generation. The SAM toolbox is publicly available and under continuous development, providing researchers with a powerful tool for validating and refining HRTFs. The latest version of the code can be accessed at: \href{https://github.com/Katarina-Poole/Spatial-Audio-Metrics}{https://github.com/Katarina-Poole/Spatial-Audio-Metrics}.

\section{Conclusion}
This work extends the SONICOM HRTF dataset with additional subjects and introduces a large set of synthetic HRTFs generated from the processed 3D scans using Mesh2HRTF. Additionally, the Spatial Audio Metrics toolbox has been developed and released to facilitate standardised HRTF evaluation and visualisation. By leveraging computational HRTF generation, this dataset significantly expands the potential for machine-learning applications, spatial audio research, and personalised HRTF synthesis. Future work will focus on increasing the number of subjects and processed scans, as well as testing the perceptual efficacy of the synthetic HRTFs within this dataset in comparison to measured HRTFs. 

\section{Acknowledgments}
This research is supported by the SONICOM project (EU Horizon 2020 RIA grant agreement ID: 101017743). We thank Xianghao Wang and Vihan Gemawat for their work on processing the 3D scans.  
\bibliography{kp_lib}

\end{document}